\documentclass[preprint,12pt]{elsarticle}
\usepackage{xcolor}

\usepackage{amssymb,bm}
\usepackage{amsthm}
\usepackage{amsmath}

\usepackage{hyperref}
\biboptions{numbers,sort&compress}

\journal{XXX}

\begin{document}

\begin{frontmatter}


\title{A scale-based study of the Reynolds number scaling for the near-wall streamwise turbulence intensity in wall turbulence}

\author[affli1]{Cheng Cheng}
\author[affli1,affli2,affli3]{Lin Fu\corref{label2}}
\cortext[label2]{Corresponding Author.}
\ead{linfu@ust.hk}
\address[affli1]{Department of Mechanical and Aerospace Engineering, The Hong Kong University of Science and Technology, Clear Water Bay, Kowloon, Hong Kong}
\address[affli2]{Department of Mathematics, The Hong Kong University of Science and Technology, Clear Water Bay, Kowloon, Hong Kong}
\address[affli3]{HKUST Shenzhen-Hong Kong Collaborative Innovation Research Institute, Futian, Shenzhen, China}

\begin{abstract}
Very recently, a defect model which depicts the growth tendency of the near-wall peak of the streamwise turbulence intensity has been developed (Chen $\&$ Sreenivasan, J. Fluid Mech. (2021), vol.908, R3).
Based on the finiteness of the near-wall turbulence production, this model predicts that the magnitude of the peak will approach a finite limit as the Reynolds number increases. 
In the present study, we revisit the basic hypotheses of the model, such as the balance between the
turbulence production and the wall dissipation in the region of peak production, the negligible effects of the logarithmic motions on the wall dissipation, and the typical time-scale that the outer-layer flow imposes on the inner layer. 
Our analyses show that some of them are not consistent with the characteristics of the wall-bounded turbulence. Moreover, based on the spectral stochastic estimation, we develop a framework to assess the wall dissipation contributed by the energy-containing eddies populating the logarithmic region, and uncover the linkage between its magnitude and the local Reynolds number.
Our results demonstrate that these multi-scale eddies make a non-negligible contribution to the formation of the wall dissipation. Based on these observations, we verify that the classical logarithmic model, which suggests a logarithmic growth of the near-wall peak of the streamwise turbulence intensity with regard to the friction Reynolds number, is more physically consistent, and still holds even with the latest high-Reynolds-number database. 
\end{abstract}



\begin{keyword}



\end{keyword}

\end{frontmatter}


\section{Introduction}
\label{sec:intro}
In recent years, in terms of the wall-bounded turbulence, the Reynolds number dependence of the peak of the streamwise turbulence intensity $\overline{u'^{2}_p}^+$ in the vicinity of the wall attracts lots of attention (here, superscript $+$ denotes the normalization with wall units, and overbar indicates the ensemble average). This phenomenon has not only been verified by direct numerical simulations (DNS), for example, in \cite{Hoyas2006}, \cite{Sillero2013}, \cite{Lee2015}, \cite{Pirozzoli2021}, but also laboratory experiments, such as in \cite{Hultmark2012}, \cite{Marusic2015}, \cite{Samie2018}, to name a few. In general, at least for the wall-bounded flow with $Re_{\tau}\le35000$ ($Re_{\tau}=hu_{\tau}/\nu$, where $h$ denotes the channel half-height, the boundary layer thickness, or the pipe radius, $u_{\tau}$ the wall friction velocity, and $\nu$ the kinematic viscosity), its establishment is evident \cite{Marusic2017}. This observation suggests the failure of the wall scaling for turbulence intensity, and indicates the scale interactions between the near-wall and the outer-layer motions. Throughout this paper, the terms `eddy' and `motion' are exchangeable.

Over the past decades, researchers have devoted themselves to understanding and predicting the growth tendency of $\overline{u'^{2}_p}^+$. 
The most influential speculation is the logarithmic model (LM), which takes the form of
\begin{equation}\label{log}
\overline{u'^{2}_p}^+=A\ln(Re_{\tau})+B,
\end{equation}
where $A$ and $B$ are two constants. Marusic et al. \cite{Marusic2017} fitted the existing DNS and experimental data, and estimated their values as $A=0.63$ and $B=0.38$. 
They further pointed out that the logarithmic growth tendency to $Re_{\tau}$ is consistent with the celebrated attached-eddy model (AEM) \cite{Townsend1976,Perry1982}. 
The increase of $\overline{u'^{2}_p}^+$ can be interpreted as the outcome of the footprints of the energy-containing eddies populating the logarithmic and outer layers on the near-wall region. It should be noted that, if Eq.~(\ref{log}) is valid, it indicates that the magnitude of $\overline{u'^{2}_p}^+$ is unbounded as the growth of $Re_{\tau}$. 

Very recently, Chen and Sreenivasan \cite{Chen2021a} developed an alternative formula to describe the growth tendency of $\overline{u'^{2}_p}^+$, and predicted that its magnitude is finite when $Re_{\tau}\to\infty$. Their derivation can be briefly introduced as follows. 
The balance equation of the streamwise turbulence kinetic energy $\frac{1}{2}\overline{u'^{2}}^+$ can be expressed as
\begin{equation}\label{balan}
0=P^++D^+-\epsilon^{+}+\Pi_x^++T_{t,x}^+,
\end{equation}
where $P^+$ is the turbulence production term, and equals to $-\overline{u^{\prime} v^{\prime}}^+\partial \bar{u}^{+}/\partial y^{+}$, i.e., the product of the mean shear and the Reynolds shear stress, $D^+=\frac{1}{2}\left(\partial^{2} / \partial y^{+2}\right) \overline{u^{\prime 2}}^{+}$ is the turbulence diffusion term, $\epsilon^{+}=\overline{\left|\nabla u^{\prime}\right|^{2+}}$ is the dissipation term, $\Pi_x^+=\overline{p^{'+}\partial u^{'+} / \partial x^+}$ is the pressure-strain term, and $T_{t,x}^+$ is the turbulence transport term. Here, $x$, $y$, and $z$ denote the coordinates in the streamwise, wall-normal, and spanwise directions, respectively. In the vicinity of the wall, the turbulence diffusion ${D}^{+}$ is thought to be balanced with the dissipation $\epsilon^+$. By the Taylor expansion of $\overline{u'^{2}}^+$, the  order of the peak magnitude can be evaluated as
\begin{equation}\label{balance}
\overline{u'^{2}_p}^+ \sim D_{w}^{+} y_{p}^{+2}=\epsilon_{w}^{+} y_{p}^{+2},
\end{equation}
where $y_{p}^{+}$ is the wall-normal position of the peak, which approximately equals 15, and the subscript $w$ stands for the value at the wall. 
If $y_{p}^{+}$ is independent on the Reynolds number, Eq.~(\ref{balance}) demonstrates that the peak value is determined by the wall dissipation. If $\epsilon_{w}^{+}$ is finite when $Re_{\tau}\to\infty$, so does $\overline{u'^{2}_p}^+$. 
They further argued that the magnitude of $\epsilon_{w}^{+}$ is bounded, since the maximum value of the turbulent production is bounded by $1/4$, and the energy production balances with the dissipation at the location of the peak production (denoted as $y_{pp}^+$, approximately equals 12). Hence, the magnitude of $\overline{u'^{2}_p}^+$ should also be bounded.

The scaling law of $\epsilon_{w}^{+}$ is then derived. Chen and Sreenivasan \cite{Chen2021a} hypothesized that the defect dissipation, i.e, $\epsilon_{d}^{+}=1/4-\epsilon_{w}^{+}$, results from the fact that a fraction of 
energy is transferred to the outer region and not dissipated in the near-wall region locally. The time-scale of this process is $\eta_o/u_{\tau}$, which
can also be considered as the superposition time-scale of outer motions on the inner layer. Here, $\eta_o=\nu^{3/4}/\epsilon_{o}^{1/4}$ is the Kolmogorov scale of the outer layer with the corresponding  dissipation $\epsilon_{o}$ as $u_{\tau}^3/h$. 
Further normalization with the wall scaling shows that $\epsilon_{d}^{+}\sim Re_{\tau}^{-1/4}$, and therefore, the wall dissipation follows that
\begin{equation}\label{ew}
\epsilon_{w}^{+}=1/4-\beta/Re_{\tau}^{1/4}, 
\end{equation}
where $\beta$ is a constant. Substituting Eq.~(\ref{ew}) into Eq.~(\ref{balance}), one can finally obtain the predition of $\overline{u'^{2}_p}^+$, which is
\begin{equation}\label{up}
\overline{u'^{2}_p}^+=\alpha(1/4-\beta/Re_{\tau}^{1/4}),
\end{equation}
where $\alpha$ is another constant. $\alpha$ and $\beta$ are fitted to be 46 and 0.42, respectively. Hereafter, we denote the defect models (\ref{ew}) and (\ref{up}) as DM. Monkewitz \cite{monkewitz2022} developed a composite model for the profiles of $\overline{u'^{2}}^+$ based on DM. Chen and Sreenivasan \cite{Chen2022} further generalized the defect model to other near-wall physics, such as the turbulence diffusion. 

Considering the two models mentioned above, some fundamental questions can be raised. First of all, which model is more physically consistent ? Second, whether we can develop a methodology to validate the LM or DM ? The present study mainly addresses these two questions. 

\section{Analyses and discussions on DM}
\begin{figure} 
	\centering  
		\includegraphics[width=1.0\linewidth]{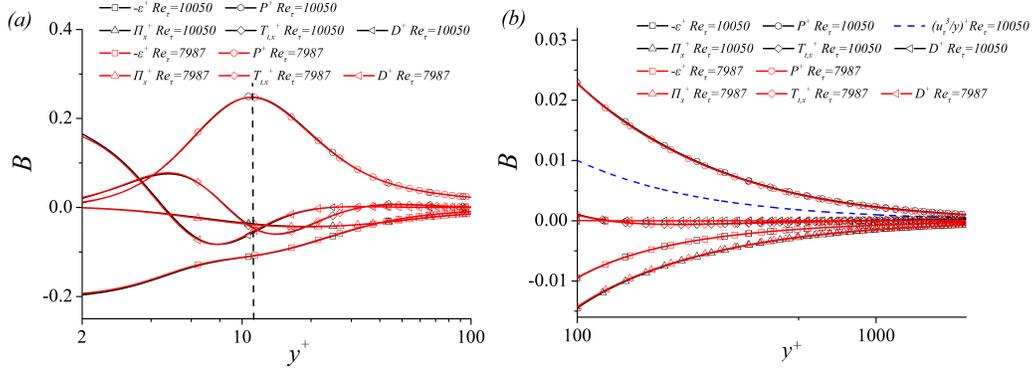} 
	\caption{Energy budget for $\frac{1}{2}\overline{u'^{2}}^+$ in channel flows at $Re_{\tau}=10050$ \cite{Hoyas2022} and $Re_{\tau}=7987$ \cite{Kaneda2021}: ($a$) near-wall region; ($b$) logarithmic region. The black dashed line in ($a$) is plotted to highlight the position $y^+_{pp}$.}
	\label{fig:balance1} 
\end{figure}

Now, we pay attention to the arguments made by Chen and Sreenivasan \cite{Chen2021a}. The first point is that the Taylor series expansion of $\overline{u'^{2}}^+$, namely Eq.~(\ref{balance}), is the result of the balance between the wall diffusion and dissipation in the vicinity of the wall. Smits et al.
\cite{Smits2021} observed that at $y^+=10$, the dominated mechanism of $\frac{1}{2}\overline{u'^{2}}^+$ is the turbulence production, and the signs of ${D}^{+}$ and $\epsilon^{+}$ are inverse at this position in a channel flow with $Re_{\tau}=5200$. Accordingly, the validity of the Taylor expansion of $\overline{u'^{2}}^+$ (Eq.~(\ref{balance})) at $y_p^+$ is ambiguous. It is noted that the similar tendency can be observed in the up-to-date DNS of channel flows at $Re_{\tau}=10050$ \cite{Hoyas2022} and $Re_{\tau}=7987$ \cite{Kaneda2021} (see Fig.~\ref{fig:balance1}($a$)). However, Smits et al. \cite{Smits2021} also found that the near-wall profiles of $\overline{u'^{2}}^+$  extracted from the existing DNS data of different types of wall turbulence collapse well if scaled with the streamwise wall-shear stress fluctuation $\overline{\tau_x^{'2}}^+$, which is identical to the wall dissipation $\epsilon_{w}^{+}$. Hence, Eq.~(\ref{balance}) is still a reasonable proposition. 
On the other side, Pirozzoli et al. \cite{Pirozzoli2021} reported that the value of $y_p^+$ in DNS of the pipe flows with $Re_{\tau}$ ranging from $180-6020$ slightly increases with $Re_{\tau}$, which may have non-negligible effects on the magnitude of $\overline{u'^{2}_p}^+$ as per Eq.~(\ref{balance}). As discussed by Chen and Sreenivasan \cite{Chen2021a}, the existing data can not offer a sufficient evidence that $y_p^+$ is a strong $Re_{\tau}$-dependent variable. For example, for a turbulent channel flow with $Re_{\tau}=180$, the peak value of $\overline{u'^{2}_p}^+$ locates at $15.3$ \cite{Moser1999}. The latest DNS data with $Re_{\tau}=10050$ also shows that $y_p^+=15.4$ \cite{Hoyas2022}. It can be seen that $y_p^+$ is insensitive to the Reynolds number. Besides, the measurement of an atmospheric boundary layer with $Re_{\tau}\approx900000$ also shows that $y_p^+\approx13.4$ with uncertainty between 11.6 and 15.2 \cite{Metzger2001}. \textcolor{black}{Schlatter and {\"O}rl{\"u} \cite{Schlatter2010} summarized the values of $y_p^+$ for different simulations of boundary layers and channels. They claimed that there exists a fairly large spread of the data with a variation of up to $10\%$ at a fixed $Re_{\tau}$ (see Fig.~4($b$) of their paper).}
Hence, in the present study, we still treat it as a Reynolds-number insensitive quantity.

\begin{figure} 
	\centering 
		\includegraphics[width=1.0\linewidth]{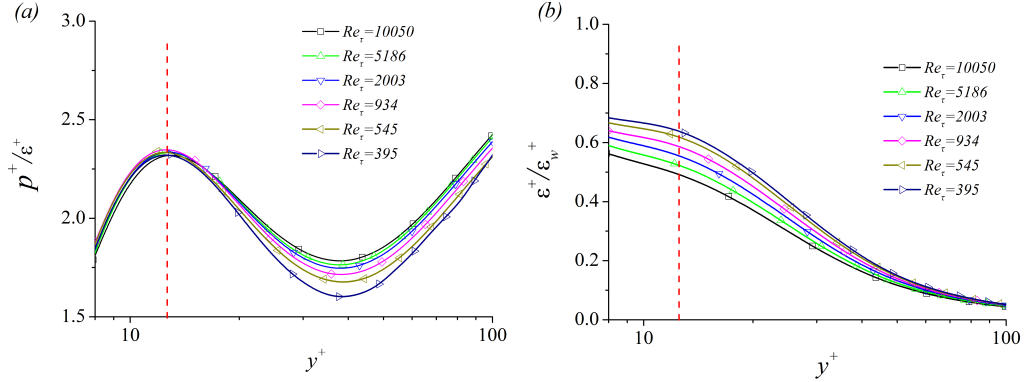} 
	\caption{Profiles of ($a$) $P^+/\epsilon^+$, and ($b$) $\epsilon^+/\epsilon_{w}^{+}$ from DNS of
		channel flows. The red dashed lines in ($a$) and ($b$) are plotted to highlight the position $y^+_{pp}$. }
	\label{fig:pk} 
\end{figure}

Another important point of DM is the defect model of $\epsilon_{w}^{+}$. 
It is established based on the hypothesis that the energy production balances with the dissipation at $y_{pp}^+$, so the wall dissipation is also bounded due to the bounded production. However, Fig.~\ref{fig:balance1}($a$) shows that the energy balance at $y_{pp}^+$ is relatively complicated, as all the terms are non-negligible. Thus, the proposition of DM that $P^+$ balances $\epsilon^+$ at the location, where the production is maximum, is questionable.
Moreover, Fig.~\ref{fig:pk}($a$) shows the ratio between $P^+$ and $\epsilon^{+}$ in the near-wall region of turbulent channel flows calculated from DNS data with $Re_{\tau}$ ranging from $O(10^2)$ to $O(10^4)$ \cite{Moser1999,DelAlamo2003,DelAlamo2006,Hoyas2006,Lee2015,Hoyas2022}. 
It can be seen that $P^+/\epsilon^{+}$ is Reynolds-number independent at $y_{pp}^+$, which can be expressed as
\begin{equation}\label{pk}
\frac{P^+(y_{pp}^+)}{\epsilon^{+}(y_{pp}^+)}=\phi,
\end{equation}
where $\phi$ is a constant, and approximately equals 2.3. Eq.~(\ref{pk}) indicates that $\epsilon^{+}(y_{pp}^+)$ is bounded on the premise of  the finiteness of $P^+(y_{pp}^+)$, and the energy produced at $y_{pp}^+$ can not be dissipated locally and completely even when $Re_{\tau}\to\infty$, which is not consistent with the deduction of DM. As DM predicts that the near-wall production will be dissipated locally and entirely when $Re_{\tau}$ is high enough.
To illustrate the relationship between $\epsilon_{w}^{+}$ and $\epsilon^{+}$, Fig.~\ref{fig:pk}($b$) shows the variations of $\epsilon^{+}/\epsilon_{w}^{+}$ as
functions of $y^+$. At $y_{pp}^+$, clear Reynolds-number dependence can be observed, which can be expressed as 
\begin{equation}\label{pk2}
\frac{\epsilon^{+}(y_{pp}^+)}{\epsilon_{w}^{+}}=f(Re_{\tau}),
\end{equation}
where $f(Re_{\tau})$ is a function of $Re_{\tau}$. The substitution of Eq.~(\ref{pk2}) into Eq.~(\ref{pk}) yields
\begin{equation}\label{pk3}
\epsilon_{w}^{+}=\frac{P^+(y_{pp}^+)}{\phi f(Re_{\tau})}.
\end{equation}
Eq.~(\ref{pk3}) suggests that the finiteness of $\epsilon_{w}^{+}$ is not only dependent on the asymptotic behavior of $P^+(y_{pp}^+)$, but also the properties of the function $f(Re_{\tau})$, which can not be determined by the existing limited database. Hence, the bounded $P^+(y_{pp}^+)$ and $\epsilon^{+}(y_{pp}^+)$ do not directly imply the boundness of $\epsilon_{w}^{+}$, thus $\overline{u'^{2}_p}^+$.

\begin{figure} 
	\centering 
		\includegraphics[width=1.0\linewidth]{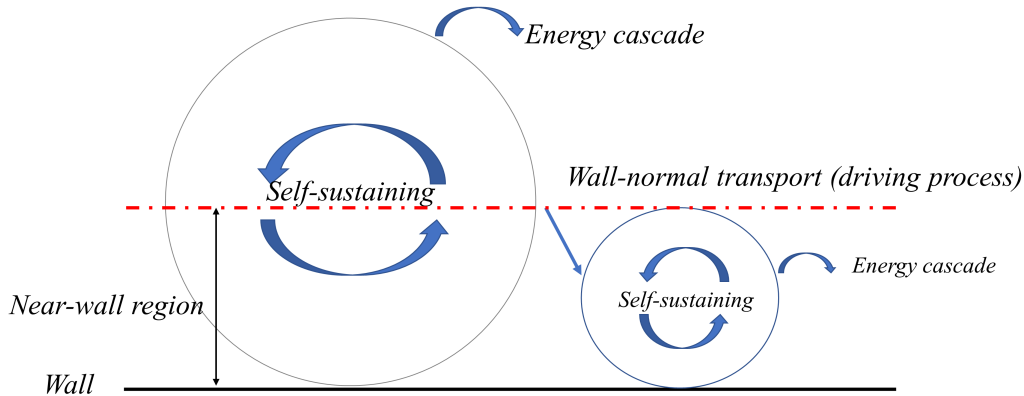} 
	\caption{A schematic of the scale interactions between the energy-containing eddies in the near-wall and the logarithmic regions. Each circle represents an individual eddy.}
	\label{fig:AEH} 
\end{figure}

The last important point of DM is that $\epsilon_{d}^{+}$ is manifested as the consequence of the near-wall energy transferred to the outer layer and dissipated nonlocally with the time-scale $\eta_o/u_{\tau}$ and with the outer dissipation rate $\epsilon_{o}$ equaling $u_{\tau}^3/h$. This scenario can be reinterpreted from the perspective of the energy-containing motions, i.e., the deviation of the wall dissipation from 1/4 can be regarded as the influences from large-scale motions (LSMs) populating the outer layer with the characteristic scale $h$, and the typical time-scale of this process is $\eta_o/u_{\tau}$. That is to say, the logarithmic region has negligible
effects on the magnitude of the wall dissipation. However, it is known to all that there are multi-scale motions in the high-Reynolds number wall-bounded turbulence \cite{Jimenez2018,Marusic2019}. In addition to the near-wall motions with the characteristic length scale $\nu/u_{\tau}$ and the LSMs with outer-flow scaling $h$, there
is still a crowd of energy-containing motions with typical length scale $y$, namely their distance from the wall, populating the logarithmic region. These $y-$scaling motions have been demonstrated to be responsible for the formation of the classical log law of the streamwise mean velocity \cite{Hwang2019}, and become more and more in number with increasing Reynolds number \cite{Marusic2019,Cheng2019}. Fig.~\ref{fig:balance1}($b$) shows the streamwise kinetic energy budget in the logarithmic region of channel flows at $Re_{\tau}=10050$ \cite{Hoyas2022} and $Re_{\tau}=7987$ \cite{Kaneda2021}. 
In the logarithmic region, $P^+$ and $\epsilon^{+}$ are not fully balanced with each other. Aside from  $\epsilon^{+}$, the pressure-strain term $\Pi_x^+$ also balances a fraction of the turbulence production. It redistributes the streamwise kinetic energy to the spanwise and wall-normal components, as the spectral scale properties of the wall-normal and spanwise pressure-strain terms are in accordance with $\Pi_x^+$ and only opposite in sign (the reader can refer to \cite{Cho2018} and \cite{Lee2019} for details). Further to this, the induced wall-normal transport would stimulate the near-wall eddies by scale interactions (see Fig.~\ref{fig:AEH}). 
Interestingly, this type of interaction process has recently been observed by Doohan et al. \cite{Doohan2021} in their numerical experiment, and named as `driving process'. 
The main results of this process are reported to be the transient amplifications of the local small-scale productions and their subsequent dissipation \cite{Doohan2021}. 
They also found that the time-scale of the driving mechanism is determined by the self-sustaining process of large-scale eddies. It can be envisioned that the time-scale is proportional to $y/u_{\tau}$, as $y$ and $u_{\tau}$ are the characteristic length scale and velocity scale of a $y-$scaling eddy populating the logarithmic region, respectively. \textcolor{black}{Our proposition is  in line with the original claim of Townsend \cite{Townsend1976}, who pointed out that the dissipation length scale in the constant-stress equilibrium layer must be proportional to the wall distance. In fact, this property is the essential building block of the celebrated attached-eddy model, because it signifies that there must exist eddies populating the logarithmic region and extending to the wall \cite{Townsend1976}.}
The study of  Lozano-Dur{\'a}n
and Jim{\'e}nez \cite{Lozano-Duran2014a} can also verify our assessment. They found that the lifetimes of the wall-attached structures extending to the logarithmic region are in direct proportion to their wall-normal heights. Hence, the corresponding dissipation imposed on the near-wall flow by the turbulence at the location $y$ can also be estimated as 
\begin{equation}\label{pk5}
\int_{y}^{h}(y^*)^{-1}u_{\tau}^3/y^*dy^{*}\sim u_{\tau}^3/y,
\end{equation}
whose variation at $Re_{\tau}=10050$ is shown in Fig.~\ref{fig:balance1}($b$). Here, $(y^*)^{-1}$ is the probability density of the $y^*-$scaling eddies \cite{Townsend1976,Hwang2018,Cheng2020}. It can be seen that it shares a similar order of magnitude with $\Pi_x^+$, which verifies the above arguments indirectly. 
Accordingly, we may conclude that the original DM is not consistent with the multi-scale characteristics of high-Reynolds number wall turbulence. 

Finally, it is worth mentioning that the physical interpretation of the typical time-scale of outer-region additional dissipation in DM, i.e., $t_a=\eta_o/u_{\tau}$, is ambiguous, since $\eta_o$ is the Kolmogorov scale in the outer region, whereas $u_{\tau}$ is the characteristic velocity scale of the energy-containing eddies.
Actually, the Kolmogorov velocity scale in outer region is $u_{\eta}=(\nu\epsilon_o)^{1/4}$, thus, the ratio between $t_a$ and Kolmogorov time-scale $t_{\eta}$ can be estimated as 
\begin{equation}\label{pk4}
\frac{t_a}{t_{\eta}}=\frac{u_{\eta}}{u_{\tau}}=(\frac{\nu}{u_{\tau}h})^{1/4}=Re_{\tau}^{-1/4}.
\end{equation}
Eq.~(\ref{pk4}) shows that $t_a$ will be far less than $t_{\eta}$ when $Re_{\tau}\to\infty$, and becomes the smallest time-scale in the outer region. This conclusion is counter-intuitive. 

\section{LM versus DM}
\subsection{A scale-based model}
According to above discussions, we can see that the defect model of wall dissipation is questionable, and the eddies occupying the logarithmic region also have impacts on the near-wall dissipation. Even if we insist that the defect model of wall dissipation is valid, and adjust the time-scale to $y/u_{\tau}$, the normalized defect dissipation $\epsilon_{d}$ at a given wall-normal height $y$ can be assessed as
\begin{equation}\label{kd2}
\epsilon_{d}^+(y^+)\sim(u_{\tau}^3/y)/(u_{\tau}^4/\nu)=(y^+)^{-1}.
\end{equation}
Thus, beyond the near-wall region, the total defect dissipation $\epsilon_{d,t}^+$  is given by
\begin{equation}\label{kd3}
\epsilon_{d,t}^+\sim \int_{y_s^+}^{h^+}(y^+)^{-1}dy^+\sim \ln(Re_{\tau}),
\end{equation}
where $y_s^+$ denotes the starting location of the logarithmic layer, which is   \textcolor{black}{classically} believed to be  $80 \leq y_s^+\leq 100$ \cite{Jimenez2018,Hu2020}. It is noted that the effects of LSMs are also included in Eq.~(\ref{kd3}). When $Re_{\tau}\to\infty$, $\epsilon_{w}^{+}$ follows a new law as
\begin{equation}\label{kd4}
\epsilon_{w}^{+}=1/4-C_1\ln(Re_{\tau}),
\end{equation}
where $C_1$ is a constant. Eq.~(\ref{kd4}) indicates that when $Re_{\tau}$ is high enough, $\epsilon_{w}^{+}$ will be negative, and, apparently, this is not possible. \textcolor{black}{Over the past two decades, several works
also reported that the lower bound of the logarithmic region is Reynolds-number-dependent, namely, $y_s^+=A_1Re_{\tau}^{0.5}$, where $A_1$ is a constant estimated between 1 and 3 \cite{Eyink2008,Marusic2013,Chin2014,Hwang2019}. This is also in line with the mesolayer scaling proposed by  Afzal \cite{Afzal1982}. For this circumstance, the total defect dissipation $\epsilon_{d,t}^+$ can  be deduced to follow Eq.~(\ref{kd3}), and thus $\epsilon_{w}^{+}$ also obeys the unphysical scaling Eq.~(\ref{kd4}).} 

Hence, it is instructive to abandon the defect description. Instead, the wall dissipation should be regarded as the accumulated outcome of the energy-containing motions, i.e., 
\begin{equation}\label{kd5}
\epsilon_{w}^{+}\sim\int_{y_s^+}^{h^+}(y^+)^{-1}dy^+=C_2\ln(Re_{\tau})+C_3,
\end{equation}
where $C_2$ and $C_3$ are two constants, and this model includes the effects of the near-wall turbulence on the wall dissipation. The substitution of Eq.~(\ref{kd5}) into Eq.~(\ref{balance}) yields the LM, i.e., Eq.~(\ref{log}). This result highlights that the LM is more physically consistent.

Upon the analyses above, it is transparent that the dispute between the two models is whether the logarithmic region has remarkable effects on the magnitude of the wall dissipation. DM regards the near-wall region as the energy source, and the departure of the wall dissipation from the ultimate state is ascribed to the energy transferred to the outer region and dissipated locally. While LM treats the generation of the wall dissipation as the additive effects resulting from the multi-scale motions. Accordingly, if we can develop a methodology to examine the effects of the log-region flow on the magnitude of wall dissipation, then we can further decide which model is physically consistent. The following subsection is focused on this issue.

\subsection{Validation by a numerical framework}\label{SP}

In this section, we employ the spectral stochastic estimation, a numerical approach, to verify our argument by analyzing the DNS database. The database used in the present study includes the incompressible turbulent channel flows at $Re_{\tau}=2003$ and $Re_{\tau}=4179$, which have been extensively validated by previous studies \cite{Hoyas2006,Jimenez2008,Lozano-Duran2014,Motoori2021,Cheng2022a}. These two simulations were conducted in a computational domain of $8\pi h\times3\pi h\times2 h$ and $2\pi h\times\pi h\times2 h$ in the streamwise, spanwise, and wall-normal directions, respectively. Forty raw snapshots are used for the ensemble average in the present study for both cases. All these data are provided by the Polytechnic University of Madrid. It is worth noting that the key results reported below are not sensitive to the  number of instantaneous flow fields employed for accumulating statistics (see Appendix A).

According to the inner-outer interaction model \cite{Marusic2010}, the large-scale motions would exert the footprints on the near-wall region, i.e., the superposition effects. Baars et al.\cite{Baars2016} demonstrated that this component (denoted as $u_{L}^{'+}(y^+)$) can be obtained by the spectral stochastic estimation of the streamwise velocity fluctuation at the logarithmic region $y_o^+$, namely, 
\begin{equation}
u_{L}^{'+}\left(x^{+}, y^{+}, z^{+}\right)=F_{x}^{-1}\left\{H_{L}\left(\lambda_{x}^{+}, y^{+}\right) F_{x}\left[u_{o}^{'+}\left(x^{+}, y_{o}^{+}, z^{+}\right)\right]\right\},
\end{equation}
where $u_{o}^{'+}$ denotes the streamwise velocity fluctuation at $y_o^+$ in the logarithmic region, and, $F_x$ and $F_x^{-1}$ denote FFT and inverse FFT in the streamwise direction, respectively. $H_L$ is the transfer kernel, which evaluates the correlation between $u^{'+}(y^+)$ and $u_{o}^{'+}(y_o^+)$ at a given length scale $\lambda_{x}^{+}$, and can be calculated as
\begin{equation}\label{HL}
H_{L}\left(\lambda_{x}^{+}, y_o^{+}\right)=\frac{\left\langle\hat{u'}\left(\lambda_{x}^{+}, y^{+}, z^{+}\right) \hat{u_o'}^{*}\left(\lambda_{x}^{+}, y_{o}^{+}, z^{+}\right)\right\rangle}{\left\langle\hat{u_o'}\left(\lambda_{x}^{+}, y_{o}^{+}, z^{+}\right) \hat{u_o'}^{*}\left(\lambda_{x}^{+}, y_{o}^{+}, z^{+}\right)\right\rangle},
\end{equation}
where $<\cdot>$ represents the averaging in the temporal and spatially homogeneous directions, $\hat{u'}$ is the Fourier
coefficient of $u'$, and $\hat{u'}^{*}$ is the complex conjugate of $\hat{u'}$. 

In this work, we mainly focus on the quantity $\epsilon_{w}^{+}$ generated by the logarithmic eddies. Thus, the predicted near-wall position $y^+$ is fixed at $y^+=0.3$, and the outer reference height $y_o^+$ varies from $100$ (namely $y_s^+$) to $0.2h^+$, i.e., the upper boundary of logarithmic region \cite{Jimenez2018,Wang2021}. Once $u_L^{'+}$ is obtained, the superposition component of $\epsilon_{w}^{+}$ can be calculated by definition (i.e., $(\frac{ \partial u_L^{'+}} { \partial y^+})^2$ at the wall), and denoted as $\epsilon_{w,L}^+(y_o^+)$. According to the hierarchical energy-containing eddies in high-Reynolds number wall turbulence \cite{Marusic2019}, $\epsilon_{w,L}^+(y_o^+)$ represents the superposition contributed from the wall-coherent motions with their heights larger than $y_o^+$. Thus, \textcolor{black}{the difference value $\Delta\epsilon_{w}^+(x,y_o^+,z)=\epsilon_{w}^+(x,y^+,z)-\epsilon_{w,L}^+(x,y_o^+,z)$ can be interpreted as the cumulation of the superposition from the wall coherent eddies with their heights less than $y_o^+$ and the contribution of the near-wall small-scale motions. Detached eddies cannot contribute to it, because they cannot interact with the wall indeed.} Considering that $y_o^+$ can be interpreted as local $Re_{\tau}$, the increase of $y_o^+$ corresponds to the enlargement of $Re_{\tau}$. In this way, the Reynolds-number effects on the quantity of $\epsilon_{w}^{+}$ can be verified directly. 

\begin{figure} 
	\centering 
		\includegraphics[width=1.0\linewidth]{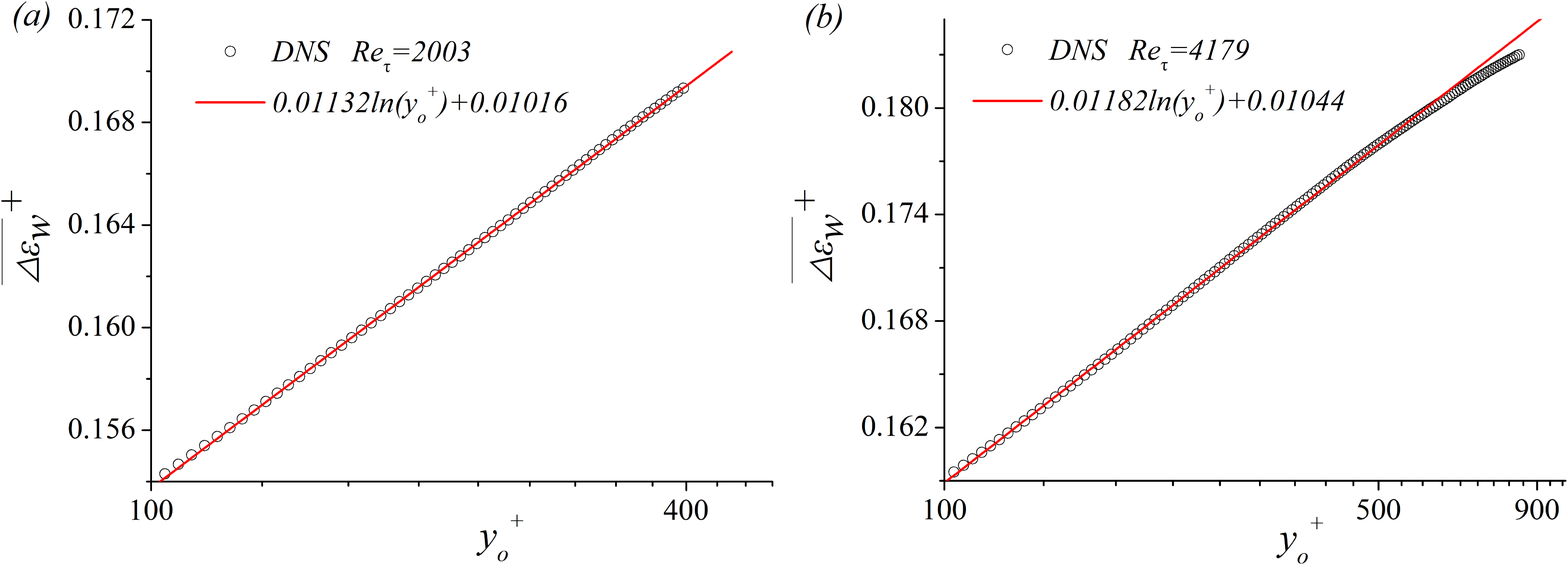} 
	\caption{Variation of the mean $\Delta\epsilon_{w}^+$ as a function of $y_o^+$ for turbulent channel flows at ($a$) $Re_{\tau}=2003$ and ($b$) $Re_{\tau}=4179$ with $y_o^+$ varying from $100$ to $0.2h^+$, and the red lines denote the corresponding exponential fitting.}
	\label{fig:AEH2} 
\end{figure}

Fig.~\ref{fig:AEH2} shows the variation of the mean $\Delta\epsilon_{w}^+$ as functions of $y_o^+$ for turbulent channel flows at $Re_{\tau}=2003$ and $Re_{\tau}=4179$.
It can be seen that $\Delta\epsilon_{w}^+$ increases with  $y_o^+$. 
It highlights the fact that the energy-containing eddies populating the logarithmic region have made a non-negligible contribution
to the formation of the wall dissipation. This essential factor, however, has not been taken into consideration by Chen and Sreenivasan \cite{Chen2021a}. Moreover, a clear exponential relationship between the mean $\Delta\epsilon_{w}^+$ and $y_o^+$ can be observed for both cases, which  provides a direct evidence for the logarithmic variation, i.e., Eq.~(\ref{kd5}). It is noted that the diversity in the upper  part of the logarithmic region of the case with $Re_{\tau}=4179$ is due to the relatively small computational domain size in this database. Again, this analysis demonstrates that the LM is more physically consistent.

The relationship between the present results and the attached-eddy model merits a discussion. The key consensus between them is that the eddies populating the logarithmic region would exert non-negligible effects on the near-wall turbulence. According to the attached-eddy hypothesis, the inactive part of the attached eddies would penetrate into the near-wall region \cite{Townsend1976,Hwang2015,Cheng2020a}. The typical time-scale $y/u_{\tau}$ discussed above can be considered as the characteristic time-scale imposing on the near-wall flow by the footprints of the attached eddies with the wall-normal height $y$. Our previous work has demonstrated that the numerical framework described above can isolate the footprints of the attached eddies populating the logarithmic region accurately \cite{Cheng2022}. Hence, the results shown in Fig.~\ref{fig:AEH2} are consistent with the attached-eddy hypothesis. Moreover, as the Reynolds number increases, the span of the logarithmic region grows with the total number of the attached eddies becoming larger simultaneously, which consequently brings about the growth of $\epsilon_w^{+}$ and $\overline{u'^{2}_p}^+$.

\section{Scalings of $\epsilon_{w}^+$ and $\overline{u'^{2}_p}^+$ in turbulent channel and pipe flows}
\begin{figure} 
	\centering 
		\includegraphics[width=1.0\linewidth]{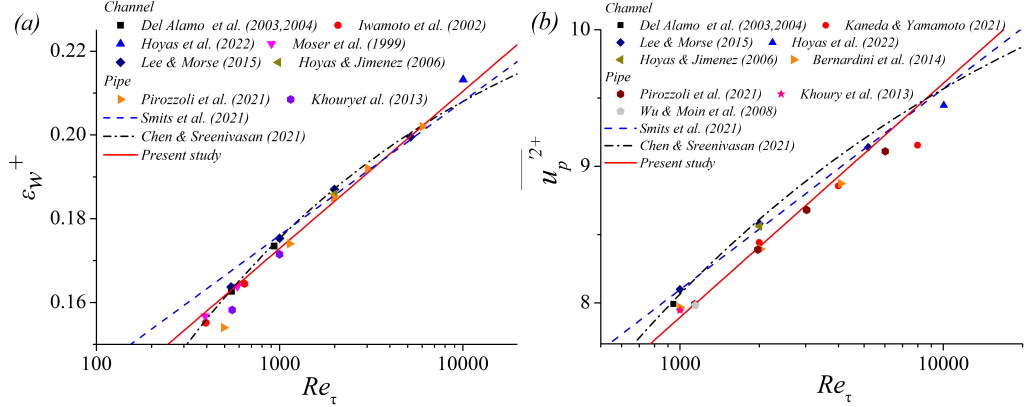} 
	\caption{Variations of ($a$) $\epsilon_{w}^+$ and ($b$) $\overline{u'^{2}_p}^+$ as functions of $Re_{\tau}$. The comprehensive DNS results of the channel and pipe flows are presented by symbols. The  blue dashed lines, the solid lines, and the dash-dotted lines denote the predictions from the logarithmic  growth fitted by Smits et al. \cite{Smits2021}, by the present study, and the defect model fitted by Chen and Sreenivasan \cite{Chen2021a}.}
	\label{fig:ddns} 
\end{figure}

Now, we focus on the direct numerical validation of $\epsilon_{w}^+$ and $\overline{u'^{2}_p}^+$ scalings with DNS database. Here, we only show the results of turbulent channel and pipe flows to highlight the differences between DM and LM, as the DNS data of channel and pipe flows are more abundant relatively, and cover a wider range of $Re_{\tau}$ \cite{Moser1999,Iwamoto2002a,DelAlamo2003,DelAlamo2004,Hoyas2006,Wu2008,ElKhoury2013,Bernardini2014,Lee2015,Kaneda2021,Pirozzoli2021,Hoyas2022}. Fig.~\ref{fig:ddns}($a$) compares the defect model of Eq.~(\ref{ew}) fitted by Chen and Sreenivasan \cite{Chen2021a}, the logarithmic growth of Eq.~(\ref{kd5}) fitted by Smits et al. \cite{Smits2021} ($\epsilon_{w}^{+}=0.08+0.0139\ln(Re_{\tau})$), and the logarithmic growth fitted by the present study for $\epsilon_{w}^+$, i.e.,
\begin{equation}\label{kd6}
\epsilon_{w}^{+}=0.0163\ln(Re_{\tau})+0.0603.
\end{equation}
The comprehensive DNS results of channel and pipe flows are included and presented by symbols. It is noted that there is a slight difference between the slope of the red lines in Fig.~\ref{fig:ddns}($a$) and Fig.~\ref{fig:AEH2}. This can be ascribed to the effects of LSMs in raw DNS data, which can not be captured by the spectral stochastic estimation conducted in section \ref{SP}. Fig.~\ref{fig:ddns}($b$)
compares the defect model of Eq.~(\ref{up}) fitted by Chen and Sreenivasan \cite{Chen2021a} and the logarithmic growth of Eq.~(\ref{log}) fitted by the present study for $\overline{u'^{2}_p}^+$ using the DNS data with $Re_{\tau}\gtrsim 1000$, i.e.,
\begin{equation}\label{kd7}
\overline{u'^{2}_p}^+=0.744\ln(Re_{\tau})+2.751=45.63 \epsilon_{w}^{+}.
\end{equation}
The similar fitting reported by \cite{Smits2021} is also included ($\overline{u'^{2}_p}^+=46(0.08+0.0139\ln(Re_{\tau}))$). Both of the two logarithmic law fittings are representative of the variational tendencies. \textcolor{black}{According to Chen and Sreenivasan \cite{Chen2021a}, a more precise version of  Eq.~(\ref{balance}) is given by}
\begin{equation}\label{kd8}
\textcolor{black}{\textcolor{black}{\overline{u'^{2}_p}^+\approx(y_p^{+2}/4)\epsilon_{w}^{+}.}}
\end{equation}
\textcolor{black}{Comparing the fitting result Eq.~(\ref{kd7}) with Eq.~(\ref{kd8}), we can obtain $y_p^+\approx13.5$. It is very close to the widely-accepted value mentioned above.}

To further quantify the fitting errors obtained with LM and DM, we calculate the root-mean-square deviation (RMSD) of the two models. It takes the form of 
\begin{equation}
\mathrm{RMSD}=\sqrt{\frac{\sum_{t=1}^{n}\left(\hat{\phi}_{t}-\phi_{t}\right)^{2}}{n}},
\end{equation}
where $n$ is the number of the DNS samples, and $\hat{\phi}_{t}$ and $\phi_{t}$ denote the predicted value and the DNS value of the target physical quantity at a given $Re_{\tau}$, respectively. Towards the databases shown in Fig.~\ref{fig:ddns}, the RMSDs of the logarithmic law fitted by the present study, by Smits et al. \cite{Smits2021}, and the defect power law proposed by Chen and Sreenivasan \cite{Chen2021a} are 0.0026, 0.0049 and 0.0029 for $\epsilon_{w}^{+}$, respectively. The counterparts of $\overline{u'^{2}_p}^+$ are 0.117, 0.135, and 0.166 for the corresponding three model predictions. It can be seen that LM is not inferior to DM in characterizing the variation tendencies of $\epsilon_{w}^{+}$ and $\overline{u'^{2}_p}^+$, as a whole.

At last, it should be acknowledged that the grid spacings of DNS can have effects on the magnitudes of the near-wall physics discussed in the present study. Yang et al.\cite{Yang2021} have shown that for resolving rare and high-intensity wall-shear stress events in wall-bounded turbulence, the desired grid spacing of DNS should be finer than the standard grid resolution. To be specific, the standard grid resolution can only resolve about $90\%$-$95\%$ rare events in a turbulent channel flow with $Re_{\tau}=10000$. In this regard, the grid spacings of the DNS conducted by Hoyas et al. \cite{Hoyas2022} with $Re_{\tau}=10050$ and Kaneda and Yamamoto  \cite{Kaneda2021} with $Re_{\tau}= 7987$ are relatively coarse (for details of the computational setups, the readers can refer to their papers). Thus, the conclusions drawn at present may not be very rigorous by just dissecting the data shown in Fig.~\ref{fig:ddns}, due to the lack of solid DNS data with higher Reynolds number. However, we can still justify which model is more consistent with the physical characteristics of the wall-bounded turbulence. Scientifically speaking, the physical implications behind these formulas, rather than the formulas themselves, are critical indeed, and they truly enhance our understandings of the complicated turbulence. After all, there is a range of functions that can match the current databases well from the point of mathematics, even better than the defect-power model and the logarithmic model discussed here with smaller RMSD, regardless of their physical significance. In summary, based on the spectral stochastic estimation, our study provides a numerical framework for investigating the physical nature of the wall-bounded turbulence. Upon the analysis of the present study, it can be concluded with caution that the classical logarithmic model is more physically consistent.

\section{Concluding remarks}
The main conclusion of the present study is that the logarithmic models, which depict the Reynolds number scalings of $\epsilon_{w}^+$ and $\overline{u'^{2}_p}^+$, are more physically consistent compared with the defect model developed by  Chen and Sreenivasan \cite{Chen2021a}. The analyses of the basic hypotheses of DM suggest that some propositions are not consistent with the characteristics of the wall turbulence. They are summarized as below.

($a$) The bounded energy production at $y_{pp}^+$ only indicates that the dissipation at $y_{pp}^+$ is finite. It does not imply that the wall dissipation is bounded.

($b$) The eddies populating the logarithmic region can exert non-trivial effects on the wall dissipation magnitude, which have not been taken into consideration by DM.

($c$) The rational time-scale with which the energy-containing eddy interacts with the near-wall flow is $y/u_{\tau}$, rather than
$\eta_o/u_{\tau}$.

Moreover, to verify our analyses, we develop a framework to quantify the wall dissipation generated by the logarithmic eddies in turbulent channel flows at $Re_{\tau}=2003$ and $Re_{\tau}=4179$. Our results show that the normalized wall dissipation stemming from these eddies has a clear exponential relationship with their wall-normal heights $y^+$. Based on the above observations, we believe that the logarithmic model is more physically consistent, and still holds even with the latest high-Reynolds-number database.

\section*{Acknowledgments}
We are grateful to the authors cited in Fig.~\ref{fig:balance1}, Fig.~\ref{fig:pk}, and Fig.~\ref{fig:ddns} for making their invaluable data available. We also thank Professor Jim\'enez for making the DNS snapshots used in section 3.2 available.
L.F. acknowledges the fund from the Research Grants Council (RGC) of the Government of Hong Kong Special Administrative Region (HKSAR) with RGC/ECS Project (No. 26200222), the fund from Guangdong Basic and Applied Basic Research Foundation (No. 2022A1515011779), and the fund from the Project of Hetao Shenzhen-Hong Kong Science and Technology Innovation Cooperation Zone (No. HZQB-KCZYB-2020083).

\section*{Appendix A. Statistic sensitivity to the number of instantaneous flow fields}\label{Ap}
\begin{figure} 
	\centering  
		\includegraphics[width=1.0\linewidth]{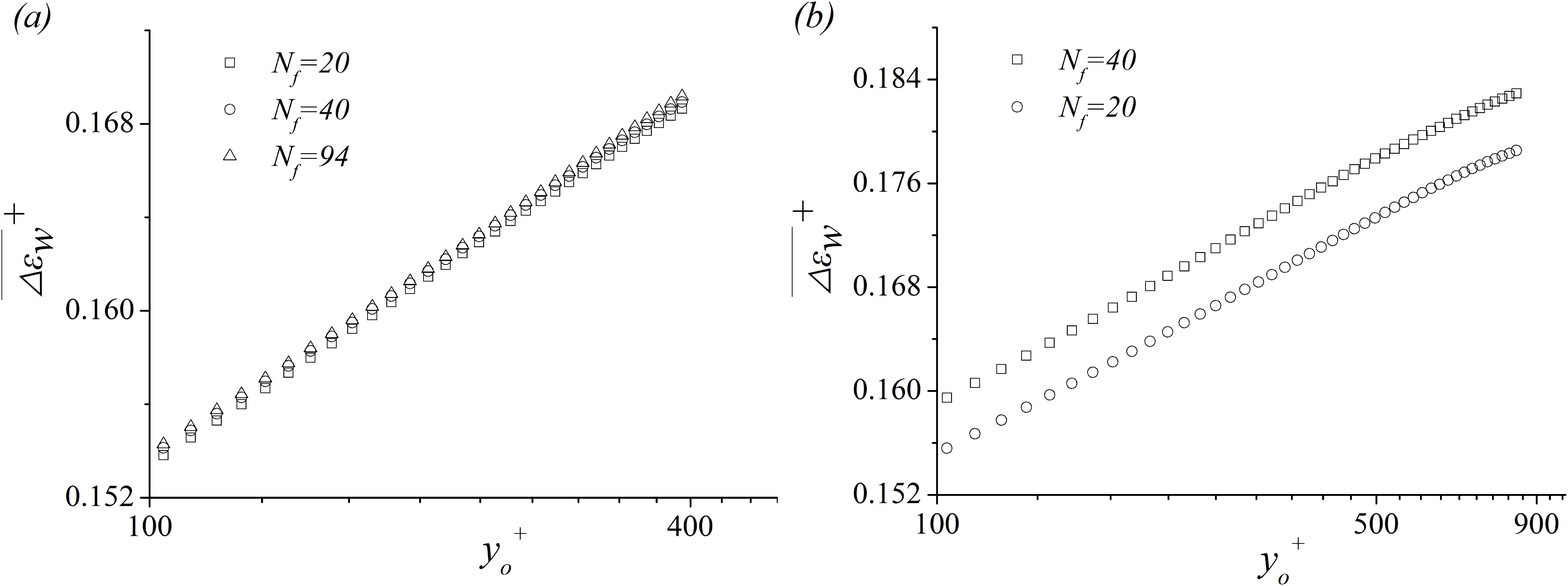} 
	\caption{Mean $\Delta\epsilon_{w}^+$ variation as functions of $y_o^+$ for the cases with ($a$) $Re_{\tau}=2003$ and ($b$) $Re_{\tau}=4179$ with different $N_f$.}
	\label{fig:less} 
\end{figure}

The influences of the number of instantaneous flow fields ($N_f$) for accumulating statistics are examined. Fig.~\ref{fig:less} shows the effects of $N_f$ on the statistic mean $\Delta\epsilon_{w}^+$ for the cases with $Re_{\tau}=2003$ and $Re_{\tau}=4179$. Alteration of the statistical samples mainly affects the intercepts of the logarithmic variations of the mean $\Delta\epsilon_{w}^+$ with respect to $y_o^+$, but the logarithmic tendencies are not changed.


\bibliographystyle{elsarticle-num} 


\end{document}